# From Declarative Languages to Declarative Processing in Computer Games[*]

Ben Sowell, Alan Demers, Johannes Gehrke, Nitin Gupta, Haoyuan Li, Walker White

{sowell, ademers, johannes, niting, haoyuan, wmwhite}@cs.cornell.edu

Cornell University
Ithaca, New York

## ABSTRACT

Recent work has shown that we can dramatically improve the performance of computer games and simulations through declarative processing: Character AI can be written in an imperative scripting language which is then compiled to relational algebra and executed by a special games engine with features similar to a main memory database system. In this paper we lay out a challenging research agenda built on these ideas.

We discuss several research ideas for novel language features to support atomic actions and reactive programming. We also explore challenges for main memory query processing in games and simulations including adaptive query plan selection, support for multi-core and parallel architectures, debugging simulation scripts, and extensions for multi-player games and virtual worlds. We believe that these research challenges will result in a dramatic change in the design of game engines over the next decade.

## 1. INTRODUCTION

Computer games are rapidly becoming an interesting area of new research [19]. In particular, scaling in games is an important problem which has largely been ignored by the research community. Games and virtual worlds have very unique workloads that differ significantly from other scalable commercial systems [16].

In the case of Massively Multiplayer Online (MMO) games, the greatest challenge to scalability is the database layer [6]. This is best illustrated by the success of EVE Online; the designers of this game focused on the database architecture first, and then built around this design. As a result, EVE Online can support a user base of 300,000 players, and over 40,000 concurrent users on a single server [6]. These numbers are an order of magnitude larger than any other MMO or virtual world.

Unfortunately, EVE Online is the exception and not the rule. Most game developers have a very difficult time in efficiently leveraging database technology in their designs. A significant part of this is the disconnect between the two processing models. In order to achieve high performance on complex queries, databases frequently perform queries and updates *set-at-a-time*. However, game developers program at the object level and design behavior for each individual object in the game. Developers use a middleware layer to translate between database and script-level object representations. This middleware is often quite fragile, which makes it very difficult to optimize and change game behavior in response to user testing [8].

In order to solve this problem, we developed a new scripting language for game designers, which we called the Scalable Games Language [17]. This language provides the performance advantages of database processing, while still allowing designers to script game components at the object level. Since the initial proposal of SGL [17], we have talked with game developers and made significant modifications to our language based on their feedback [1, 18]. Indeed, the specification of SGL is still evolving as we add new features in order to make it appealing to game developers and functionally complete.

In this paper, we discuss our current and future work on the design and implementation of the SGL language. We begin in Section 2 with a short retrospective and discuss our move toward an imperative language. We then present in Section 3 an ambitious forward-looking vision and map out what we see as the important features necessary for SGL to be embraced by the game development community at large. In Section 4, we discuss some of the implementation challenges that arise from our design of SGL. We present these as a challenge to the database research community. We end in Section 5 with some concluding remarks.

## 2. THE DESIGN OF SGL

Though game developers want to program game objects individually, the fundamental observation behind SGL is that game objects far outnumber possible game behaviors. Thus we can dramatically improve the performance of computer games and other discrete-event simulations by using database query processing and indexing technology to process these behaviors set-at-a-time [17].

---

[*]This material is based upon work supported by the AFOSR under Award No. FA9550-07-1-0437, the National Science Foundation under Grant IIS-0725260, and by a grant from Microsoft. Any opinions, findings, and conclusions or recommendations expressed in this publication are those of the authors and do not necessarily reflect the views of the sponsors.

4[th] *Biennial Conference on Innovative Data Systems Research (CIDR)*
January 4-7, 2009, Asilomar, California, USA.

SGL is a scripting language used to specify character behavior. In most modern games, the *discrete simulation engine* is responsible for executing character scripts. At every timestep, or "tick," the engine executes each script and updates the state of the world. Note that a tick is the smallest unit of time that we consider in a game. Two actions that occur during the same tick are treated as simultaneous. This will allow us to reorder computation to achieve better performance.

The primary technique used by the current version of SGL is a design pattern which we call the *state-effect pattern* [18]. In this pattern, game data is partitioned into two components, called *state variables* and *effect variables*. During each tick, the game does the following:

1. (Query Step) Query the state variables. State variables are read only.

2. (Effect Step) Compute values for effect variables. Effects are combined using aggregate functions.

3. (Update Step) Compute new values for state variables from the effect values and the previous state values.

Game developers already use this pattern for applications like particle systems. They leverage the fact that steps (1) and (2) are read-only to exploit parallelism. SGL provides compiler support for this pattern, generating relational algebra expressions from steps (1) and (2). Thus we can exploit parallelism, set-at-a-time processing, and a wide range of database query optimization techniques without any expertise from the game designer. Step (3), on the other hand, does not easily adapt itself to database processing. In order to integrate SGL with tradtional game engines, it must include non-scripted functionality, such as the physics engine or pathfinding. These pose some challenges for integrating SGL into a game engine and are discussed below.

## 2.1 Declarative Scripting without SQL

The initial version of SGL [17] was geared towards real-time strategy (RTS) games, and this led to a number of simplifying assumptions. Objects in RTS games are generally quite simple, so the language supported only a single relation (called the *environment table*) with a fixed schema. There was also no built-in support for iteration; the language included only variable assignments and conditionals. This was offset by the ability to execute arbitrary SQL from within SGL scripts, a feature which turned out to be extremely powerful. Using it, we were able to emulate most of the script-level behavior from the popular commercial game *Warcraft III* [17].

The heavy reliance on SQL required the programmer who wrote SGL scripts to have a thorough understanding of both the SQL language and the underlying set-at-a-time processing model. This turned out to be a serious drawback. While the declarative *processing* model has significant advantages for making games scalable and efficient, a declarative *programming* model was unnatural to many game programmers.[1] They want to think sequentially in terms of the sequence of observations and actions performed by individual non-player characters (NPCs) in the game. Indeed many

[1] The authors have interesting anecdotes from their experiences at game conferences to convince the game community of this first version of SGL; we will share these during our talk at the CIDR Conference.

```
class Unit {
state:
  number player = 0;
  number x = 0;
  number y = 0;
  number health = 0;
    ...
effects:
  number vx : avg;
  number vy : avg;
  number damage : sum;
    ...
}
```

**Figure 1: Class Declaration Fragment**

game developers have been quite vocal in other venues about their difficulties with SQL [11].

After this feedback on our design, we realized that it is is not necessary to expose SQL to game programmers at all. To eliminate SQL, we redesigned SGL to appear much more "object-oriented" [18]. In our new design, SQL schemas are replaced by *class definitions*, in which the distinction between state variables and effect variables is made explicit, and the aggregate function is specified for each effect variable. A fragment of an SGL class definition is shown in Figure 1.

Another advantage of this approach is that we are no longer restricted to a single environment table whose schema is known by the programmer. The SGL compiler can generate the tables from these class definitions without the programmer knowing anything about them. Indeed, this has allowed us to experiment with the best schema representation for a given class. For example, we have discovered that it is often best to break a class up into multiple tables containing those attributes that commonly appear in expressions together. In other cases it is preferable to construct a single table for all of the state variables, and a separate table for each individual effect variable.

Since the compiler generates the relational schema, we were also able to add much more complicated data types to SGL. SGL now supports reference and (unordered) set data types as either state or effect variables. These would have been difficult to support if the game designer had to explicitly design the schema in SQL. This advance in SGL is especially appealing to the developers of role-playing games (RPGs) who have a lot of container objects that force them to construct very complicated schemas [8].

In an SGL script, state variables are *read-only* while effect variables are *write-only*. To make this explicit a script may include assignments of the form

```
x <- expression
```

where `x` is an effect variable. This indicates that a value is written to `x` but that it is not immediately available for reading. At the end of each tick, all the values assigned to $x$ in this way are combined using the aggregation function associated with `x` in the class definition.

We can exploit this limited form of assignment to provide a useful form of bounded iteration. We call this iteration an `accum`-loop, and it has the following syntax:

```
accum ⟨TYPE⟩ ⟨identifier⟩₁ with ⟨COMBINATOR⟩
      over ⟨TYPE⟩ ⟨identifier⟩₂ from ⟨EXP⟩ {
    ⟨BLOCK⟩₁
} in {
    ⟨BLOCK⟩₂
}
```

```
accum number cnt with sum over unit w from UNIT {
  if (u.x >= x-range && u.x <= x+range &&
      u.y >= y-range && u.y <= y+range) {
    cnt <- 1;
  }
} in {
  ...
}
```

**Figure 2: Sample `Accum`-Loop**

Informally, this loop uses the first code block, $\langle\texttt{BLOCK}\rangle_1$, to iterate over the elements in the set $\langle\texttt{EXP}\rangle$, and then makes the results of that iteration available to the second code block, $\langle\texttt{BLOCK}\rangle_2$. Within the first code block, the variable $\langle\text{identifier}\rangle_1$ is write-only like an effect field. We make no guarantees about the order in which the accumulation loop is processed; the elements in $\langle\texttt{EXP}\rangle$ can be processed in any order, or even in parallel. Therefore, within $\langle\texttt{BLOCK}\rangle_1$, the variable $\langle\text{identifier}\rangle_1$ may never be read, and values are assigned to it using the same "`<-`" operator as effect fields.

At the completion of $\langle\texttt{BLOCK}\rangle_1$, the `accum`-loop combines all of the values assigned to $\langle\text{identifier}\rangle_1$ using the combination function $\langle\texttt{COMBINATOR}\rangle$, which can be any function allowed for effect fields. Once the values are combined, they may be read from the variable in the second code block, $\langle\texttt{BLOCK}\rangle_2$, where the variable $\langle\text{identifier}\rangle_1$ is read-only like a state field. One can think of `accum`-loops as using the state-effect pattern "locally" within a script.

The advantage of the `accum`-loop is that it allows SGL programmers to write specialized aggregate computations and apply effects to dynamically computed collections without having to rely on external SQL functions. For example, the `accum`-loop in Figure 2 counts the number of other units within a certain range. Despite the fact that this script looks imperative, it can still be compiled to a relational algebra query, so the processing model remains declarative.

We emphasize the importance of this distinction between a declarative *language* and declarative *processing*. For some applications a declarative language can help users think more clearly, but in ours (game programming) it conflicts with the way developers usually think about the application domain. The redesigned SGL is attractive to game programmers, while retaining the scalability advantages that result from compiling to relational algebra and performing set-at-a-time processing.

## 2.2 Improving the Update Step

As mentioned above, at the end of each tick state attributes are updated based on the effects computed during the tick. In the initial version of SGL these update procedures were limited to simple expressions. One example would be

```
health = health - damage;
```

which simply subtracts the value of the `damage` effect from the previous value of the `health` state attribute. This type of update is sufficient for many simple examples, but it fails to take into account the other components that are present in most commercial game engines. For instance, most games include a dedicated physics engine that examines forces and uses them to update the positions and velocities of game objects. They often include sophisticated and highly optimized collision detection and constraint solving algorithms and cannot easily be expressed in SGL [15]. Nevertheless, they need to be able to update state variables, such as position, that are used by SGL scripts. Put differently, this expression-based update model requires scripters to write explicit update rules for each attribute. In a real game, basic functionality such as movement and collision detection will be written in a low-level language as part of the engine.

There is also the issue that the output of the physics engine often does not correspond exactly to the effect assignments (or "intentions") of any individual script. For example, if two characters try to move to the same position, the physics engine may move them to adjacent locations that were not mentioned in either script. This is generally correct behavior, but it does not quite correspond to the SGL semantics described above. The physics engine takes effect assignments as input, but its actions are not expressible in SGL [15].

There are a number of other subsystems that behave like the physics engine. These include AI planning, such as pathfinding, and transaction processing (see Section 3.1). Rather than try to extend SGL to support these disparate features, we simply broaden our definition of update rules to include these more complex subsystems. In particular, each state attribute is assigned to (or owned by) a single *update component*, such as the physics engine, that updates its values at each tick. We require that the state variables be strictly partitioned among these components to avoid introducing any ordering constraints. Since the update components may produce unexpected results (e.g. the physics engine changes the target location), scripts also need to be able to determine what happened during the previous tick. We consider this problem in Section 3.2.

## 3. EXTENDING SGL

The SGL system consists of two components: the SGL language itself and the query execution engine. Each of these presents a number of very exciting research problems. In this section we discuss research directions related to the SGL language, and in Section 4 we consider the implementation.

## 3.1 Transactions and Atomicity

One limitation of the state-effect pattern is that a script cannot change or modify its actions based on observations made during the same tick. This can make it difficult to coordinate actions between several scripts. For example, many games include a financial component in which characters exchange in-game currency for special items or abilities. In general, we would like to ensure that these transactions are well-behaved. We cannot guarantee full ACID behavior in SGL, but at the very least we would like transactions to be atomic and consistent. Money should be deducted from my account only if I receive the appropriate items, and the exchange should abort if it would produce an inconsistent state such as causing my bank account to go negative. We should mention that these financial exchanges are actually quite general. NPCs frequently have attributes like health or mana (magic power), and they "spend" these resources during the course of a game, e.g., when they are injured or cast a spell. Almost any exchange of finite resources should be written as a financial exchange, so transactions will be relevant to a wide variety of common game scenarios.

To some extent, SGL already provides atomicity using effect combinators (which are formally applied by the $\oplus$ operator [17]). All writes are performed during the update phase, and conflicts are handled using effect-specific combination operations. Unfortunately, this is not sufficient to program the financial transactions described above since *all* writes succeed — even those that conflict. For example, in a single tick, one of the scripts would set effects in both objects that deducted the appropriate amount of money and transferred the items. In this case, the exchange would be atomic, but it could lead to inconsistencies if multiple characters (i.e., scripts) try to purchase the same item. Such duplication (or "duping") bugs are very common, and they can significantly impact the success of a game [7].

Another approach for transactions would involve a multi-tick protocol. Several buyers could propose to purchase an item, and the seller would choose one using $\oplus$. The actual exchange would happen during the next tick. Unfortunately, this can also lead to inconsistencies. Suppose that the seller decides to sell to buyer $b$. If $b$ is robbed during the same tick as the exchange, then he or she may end up with a negative balance.

The problem with both of these solutions is that SGL does not guarantee any type of isolation. SGL cannot avoid conflicting writes, but instead uses $\oplus$ to combine them. This is the right behavior for games — after all there is no isolation in the real world — but it causes problems when transactional semantics are required, since scripts cannot see the actions of other characters (i.e., other effect assignments) during a tick.

We solve this problem by introducing a transaction construct to the language. In SGL, a transaction is a region of code that is marked atomic, along with some constraints over state attributes. For example, we might mark the entire exchange described above as atomic and include the constraint `account > 0`. During the update rule, the game engine is then responsible for choosing a subset of the transactions issued during the tick that do not violate any constraints. The remaining transactions abort, and their effect assignments are not applied. The transaction engine fits directly into the *update component* model described in Section 2.2. It updates all the constrained state variables and may accept or reject whole transactions depending on whether they violate their respective constraints.

## 3.2 Multi-tick and Reactive Scripting

With the addition of transactions, we believe that SGL is sufficiently expressive to describe sophisticated NPCs in a wide variety of games and simulations. Unfortunately, the scripts tend to become large and unwieldy for even moderately complicated games. The problem arises because SGL scripts are essentially large state machines. If we want an NPC to move to location $(x,y)$, pick up item $i$, and then attack character $c$, then we need a state variable to keep track of the character's progress. This can become unmanageable when each script contains many complex behaviors. This problem is not unique to SGL — many game scripting languages are modeled as state machines [13] — but the problem is particularly acute for us since we are targeting much more complex behaviors than traditional game scripts. In this section we describe two constructs that we are developing to simplify the programming of complex scripts.

**Multi-Tick Scripts.** We first describe adding support for multi-tick scripts. As we mentioned above, any sequence of behaviors will require more than one tick to execute, and we have to introduce a state variable to keep track of the current step in the sequence. One way to simplify such scripts would be to maintain this state implicitly. We are currently implementing a `waitNextTick` operation that saves the state of the script and resumes at the next tick. For instance, the example described above could be written as:

```
moveX <- x; moveY <- y;  // move to (2,3)
waitNextTick;
itemsAcquired <= i;  // Pick up item i.
waitNextTick;
c.damage <- 1;  // Attack character c.
```

Note that `waitNextTick` essentially serves as a program counter, indicating where the program should resume at the next tick. We forbid `waitNextTick` inside the first block of an `accum`-loop. Since each "iteration" of the `accum`-loop happens in parallel, it does not make sense to delay in the middle. We also prohibit `waitNextTick` calls inside transactions, though we are still working to understand the exact interaction between these two constructs.

With these restrictions, there is a direct translation between multi-tick programs using `waitNextTick` and standard single-tick SGL programs. We can simply reintroduce state variables and conditions to indicate where the script should begin. Our objective here is not to fundamentally change the tick-at-time processing model of SGL, but to add syntactic features that make it more palatable to the programmer.

**Reactive Programming.** In addition to storing the program counter, scripts also use state attributes to determine what happened during the previous tick. Since a script cannot read any effects during a tick, it must read its state attributes at the beginning of a tick to determine the actions of other NPCs and information about which transactions committed. This means that most SGL scripts will begin with a large number of `if-then-else` statements that can make writing and maintaining these scripts a challenge. This problem becomes even more pronounced when we add `waitNextTick`, since the conditionals must be replicated after each such statement (i.e. at every place a tick may start).

An alternative would be to add a reactive or event-driven component to SGL. Rather than explicitly testing for every possible condition at the beginning of each tick, scripts could be notified when certain conditions are met. The simplest version of this feature would simply be syntactic sugar for the sequence of conditionals described above. Scripts could register handlers with the engine that include a condition and some effect assignments. At the end of the update phase, those handlers with conditions that evaluate to true would be executed and set some effects for the next tick. This would not necessarily reduce the amount of code that developers need to write, but it might help separate the high-level *intentional* parts of a script from the more detailed *reactive* parts.

We are currently exploring more sophisticated reactive programming models for SGL that are based on this distinction between intentions and their effects. Intentions are high-level behaviors that may last for several ticks and include calls to `waitNextTick`. As noted above, characters also need to respond to external actions on a tick-by-tick basis. The effect of these reactive programs may depend on the intention that an NPC is currently executing. For example, if an NPC has the intention to speak with some other character, he or she may want to interrupt this in order to respond to an attack. On the other hand, a character engaged in battle would likely not change his or her intention to respond to dialogue. In addition to the handlers described above, we need a mechanism to interrupt multi-tick scripts and reset the program counter.

In many ways, this notion of interruptible and resumable intentions is analogous to exception handling in a general-purpose programming language. Most modern languages use a "termination model" for exceptions, where they respond to an exception by discarding stack frames until a handler is found. However, there has also been work on resumable exceptions [12]. When a resumable exception is thrown, the runtime searches up the program stack for a handler without immediately discarding stack frames. When a handler is invoked, it has the option of resuming program execution at the point where the exception was thrown. In our setting, resumable exceptions correspond to maintaining the program counter established by `waitNextTick`.

There is clearly a tradeoff between language expressiveness and simplicity, and we are still working to find the right balance. Our goal is to develop language features that free the developer from the tedium of programming state machines without straying too far from SGL's simple imperative model.

### 3.3 Debugging

Regardless of how well a language is designed, developers need some mechanisms for debugging and reasoning about their programs. Part of this is a software engineering issue, but SGL presents several special research challenges. First of all, SGL is a data-parallel language — NPCs are processed (conceptually) in parallel. This makes many standard debugging techniques ineffective. Even the time-honored practice of debugging with print statements is of limited utility, since the same script is executed hundreds or thousands of times for different NPCs during each tick.

There has been a flurry of work recently on data-parallel languages for large distributed environments, but debugging remains largely unexplored [4, 9, 10]. In [9], the authors do describe the Pig Pen environment for debugging Yahoo's Pig-Latin language. Their approach is to produce example tuples to illustrate the output of each query. This is useful for testing long-running programs that operate on millions or billions of tuples, but SGL scripts execute in real-time and target thousands of NPCs with very complex scripts. In our experience, the depth of the query tree often means that bugs manifest themselves as empty relations, which propagate up the join tree and make examples ineffective.

Debugging SGL is also challenging because the programming and execution models are so different. Programmers think imperatively about the actions taken by a single NPC during a tick, but the runtime processes them set-at-a-time in a relational query plan. Since the optimizer also performs algebraic rewrites, there is often no direct correspondence between a point in the query plan and a point in the SGL script. We are just beginning to consider this question, but the following are some initial desiderata:

- Developers should be able to inspect the value of state attributes at tick boundaries. This is fairly easy to accomplish using a mapping between relation table names and SGL attributes.
- SGL should include support for logging, including resumable checkpoints.
- Developers should be able to select an individual NPC and view the effects assigned to it.

Note that these features require modifying both the language and the runtime, and we will need to build tools to inspect the operation of a running game. Ultimately, determining the "right" set of debugging features is a matter of gaining more experience with SGL, and we will continue to study this problem as we develop substantial games in the language.

## 4. IMPLEMENTATION ISSUES

Along with developing the SGL language, we are building an extensible game engine to demonstrate the scalability of our solutions. The core of the architecture is a main memory specialized query engine. One of the major advantages of our system is that it uses relational algebra, so we can leverage much of the existing work on query optimization. Nevertheless, there are unique opportunities for optimization that we discuss in the following sections.

### 4.1 Adaptive Query Optimization

The two dominant features of the SGL workload are that (1) the same query is executed repeatedly at every tick, and (2) we expect a large fraction of the data to change at every tick. It is likely that each NPC will move at each tick, for instance, so we may have $O(n)$ updates for $n$ characters. These updates are not random, however, and we can exploit some of their structure. For example, most NPCs will move "continuously" to a nearby location rather than jump large distances. Other attributes will behave differently and some games may include exotic features like teleportation, but in most cases games will transition periodically between a small number of different states (or workloads). A strategy game will look very different when characters are exploring than when they are fighting, but it is unlikely that the game will switch back-and-forth between the two very frequently.

This can have a dramatic affect on query optimization. The size of intermediate tables can vary dramatically between states, for instance, and this may significantly change the best join ordering. We are currently exploring the idea of compiling several query plans optimized for different workloads and to switch between them as the game progresses.

Even with a well-chosen set of query plans, we still must decide when to switch between them during the game. Ideally we would like to keep some sort of statistics about the distribution of our data, but this is difficult to do efficiently. Since many of our joins involve multi-dimensional range predicates, a histogram is not sufficient, and other approaches are poorly suited to our real-time demands [2].

### 4.2 Parallelism and Virtual Worlds

Since most PCs today ship with multiple cores, game developers are becoming increasingly eager to parallelize game engines. Because SGL compiles directly to a relational query plan, we can easily parallelize scripts using standard techniques [5, 14]. Since character data is memory-resident and fairly small, the cost of each operator is likely to be low, but we can exploit inter-operator parallelism by executing different subtrees of the query plan on different cores. Since all tables are read-only until the update phase, effect computation can occur without synchronization.

We are also considering how to optimize the SGL architecture for a shared-nothing cluster. This is particularly important for large-scale simulations, which can be orders-of-magnitude bigger than commercial games. We are currently working on a project to simulate traffic networks with millions of vehicles, and this will surely require a clustered architecture.

A cluster presents different challenges than a multicore system. Our application data are still likely to fit into the main memory of a single machine, but SGL makes extensive use of large multi-dimensional orthogonal range tree indices. Each of these trees takes $\Theta(n \log^{d-1} n)$ space, where $n$ is the number of elements in the tree and $d$ is its dimensionality [3]. Thus a tree with 100,000 entries of 16 bytes each takes about 2 GB to store. As the dimensionality and number of characters increase, this will quickly exhaust the main memory of a single machine. Thus an interesting research question is to consider techniques to partition indices across multiple nodes.

In addition to simulations, running SGL on a shared-nothing cluster is also highly relevant for massively multi-player online games (MMOs) and virtual worlds, which are often centrally hosted. Here the problem is not only to parallelize SGL, but to extend the language to abstract away network complexities such as latency, update conflicts, and rollbacks. Different games are sensitive to these parameters in different ways, and this makes the problem inherently difficult. For example, games that depend on split-second reaction times are likely very sensitive to latency, and designers may want the ability to specify certain latency requirements in the system.

## 5. CONCLUSIONS AND ONGOING WORK

In this paper we have described our ongoing work with SGL, a scripting language and engine for developing scalable games and simulations. SGL uses a declarative processing model, and it leverages many of the query optimization techniques developed by the database community over the past several decades. While SQL is the lingua franca of the database community, it is often unfamiliar or ill-suited to other domains. In games and simulations, developers think about programming the behavior of a single agent, which necessitates an imperative language. More broadly, the imperative paradigm is familiar to programmers from a wide variety of domains, and by developing the techniques to translate imperative programs to relational query plans, we hope to promote the use of declarative processing without the limitations of declarative programming.

We are currently working on a prototype of a full SGL system (including compiler, run-time system, and development environment with debugger) that we plan to use in a game design class at Cornell in the Spring of 2009. By releasing the language to students (and into the wild), we hope to learn more about how to make SGL easier to use for developers and designers who are unfamiliar with database concepts. So far usability has been the compass in our quest, and we continue to solicit feedback about our choices with game companies and at the major game conferences.

We hope that our research will not only change the way that games and simulations are programmed and processed, but that we can also help bring declarative processing techniques into the mainstream.